\documentclass[apj,twocolumn,twocolappendix,numberedappendix]{openjournal}
\usepackage{amsmath}
\usepackage{booktabs}
\usepackage{multirow}
\usepackage{color}
\usepackage{soul}
\usepackage{threeparttable}
\usepackage{float}
\usepackage{graphicx}
\usepackage{CJK}
\usepackage{xspace}
\usepackage{afterpage}
\usepackage{placeins}
\usepackage[breaklinks,colorlinks,citecolor=blue,urlcolor=blue,linkcolor=blue,filecolor=blue]{hyperref}

\shorttitle{Low dust content in a metal-rich galaxy at $z=7.13$}
\shortauthors{Heintz et al.}

\usepackage[export]{adjustbox}

\newcommand{\orcidauthor}[3]{\author{\href{http://orcid.org/#1}{#2$^{#3}$}}}

\newcommand{\cii}{C\,{\sc ii}}
\newcommand{\hi}{H\,{\sc i}}
\newcommand{\oiii}{O\,{\sc iii}}
\newcommand{\oii}{O\,{\sc ii}}
\newcommand{\jwst}{{\em JWST}}

\begin{document}

\title{\vspace{-0.8cm} Evidence for inefficient dust production in a massive, metal-rich galaxy at $z=7.13$ \\ uncovered by {\em JWST} and ALMA \vspace{-1.6cm} }

\orcidauthor{0000-0002-9389-7413}{Kasper~E.~Heintz$^*$}{1,2,3}
\orcidauthor{0000-0002-4465-8264}{Darach~Watson}{1,2}
\orcidauthor{0000-0001-6477-4011}{Francesco~Valentino}{1,4}
\orcidauthor{0000-0003-0205-9826}{Rashmi~Gottumukkala}{1,2}
\orcidauthor{0000-0002-7064-4309}{Desika~Narayanan}{5,1,2}
\orcidauthor{0000-0001-9320-4958}{Robert~M.~Yates}{6,7}
\orcidauthor{0009-0005-4175-4890}{Chamilla~Terp}{1,2}
\orcidauthor{0000-0002-7755-8649}{Negin~Nezhad}{8}
\orcidauthor{0000-0003-1614-196X}{John~R.~Weaver$^\dagger$}{9}
\orcidauthor{0000-0002-7595-121X}{Joris~Witstok}{1,2}
\orcidauthor{0000-0003-2680-005X}{Gabriel~Brammer}{1,2}
\orcidauthor{0000-0001-8169-7273}{Anja~C.~Andersen}{2}
\orcidauthor{0000-0002-5460-6126}{Albert~Sneppen}{1,2}
\orcidauthor{0009-0001-2808-4918}{Clara~L.~Pollock}{1,2}
\orcidauthor{0000-0002-4205-9567}{Hiddo Algera}{10}
\orcidauthor{0009-0009-2671-4160}{Lucie~E.~Rowland}{11}
\orcidauthor{0000-0001-5851-6649}{Pascal~A.~Oesch}{3,1,2}
\orcidauthor{0000-0002-4872-2294}{Georgios~Magdis}{1,4}
\orcidauthor{0009-0004-6791-9246}{Giorgos~Nikopoulos}{1,2}
\orcidauthor{0000-0002-7821-8873}{Kirsten~K.~Knudsen}{12}

\thanks{$^*$E-mail: \href{mailto:keheintz@nbi.ku.dk}{keheintz@nbi.ku.dk}}
\thanks{$^\dagger$Brinson Prize Fellow}

\affiliation{$^1$ Cosmic Dawn Center (DAWN), Copenhagen, Denmark}
\affiliation{$^2$ Niels Bohr Institute, University of Copenhagen, Jagtvej 155A, Copenhagen N, DK-2200, Denmark}
\affiliation{$^3$ Department of Astronomy, University of Geneva, Chemin Pegasi 51, 1290 Versoix, Switzerland}
\affiliation{$^4$ DTU Space, Technical University of Denmark, Elektrovej, Building 328, 2800 Kgs. Lyngby, Denmark}
\affiliation{$^5$ Department of Astronomy, University of Florida, 211 Bryant Space Sciences Center, Gainesville, FL 32611, USA}
\affiliation{$^6$ Centre for Astrophysics Research, University of Hertfordshire, Hatfield, AL10 9AB, UK}
\affiliation{$^7$ Department of Physics, University of Surrey, Stag Hill, Guildford, GU2 7XH, UK}
\affiliation{$^8$ Department of Physics and Astronomy, University of California, Riverside, 900 University Avenue, Riverside, CA 92521, USA}
\affiliation{$^9$ MIT Kavli Institute for Astrophysics and Space Research, 70 Vassar Street, Cambridge, MA 02139, USA}
\affiliation{$^{10}$ Institute of Astronomy and Astrophysics, Academia Sinica, 11F of Astronomy-Mathematics Building, No.1, Sec. 4, Roosevelt Rd, Taipei 106216, Taiwan, R.O.C.}
\affiliation{$^{11}$ Leiden Observatory, Leiden University, P.O. Box 9513, NL-2300 RA Leiden, the Netherlands}
\affiliation{$^{12}$ Department of Space, Earth \& Environment, Chalmers University of Technology, SE-412 96 Gothenburg, Sweden}

%\affil[1]{Cosmic Dawn Center (DAWN), Denmark}
%\affil[2]{Niels Bohr Institute, University of Copenhagen, Jagtvej 128, 2200 Copenhagen N, Denmark}
%\affil[3]{Department of Astronomy, University of Geneva, Chemin Pegasi 51, 1290 Versoix, Switzerland}
%\affil[4]{DTU Space, Technical University of Denmark, Elektrovej, Building 328, 2800 Kgs. Lyngby, Denmark}
%\affil[5]{Department of Astronomy, University of Florida, 211 Bryant Space Sciences Center, Gainesville, FL 32611, USA}
%\affil[6]{Centre for Astrophysics Research, University of Hertfordshire, Hatfield, AL10 9AB, UK}
%\affil[7]{Department of Physics, University of Surrey, Stag Hill, Guildford, GU2 7XH, UK}
%\affil[8]{Department of Physics and Astronomy, University of California, Riverside, 900 University Avenue, Riverside, CA 92521, USA}
%\affil[9]{Department of Astronomy, University of Massachusetts, Amherst, MA 01003, USA}
%\affil[10]{Institute of Astronomy and Astrophysics, Academia Sinica, 11F of Astronomy-Mathematics Building, No.1, Sec. 4, Roosevelt Rd, Taipei 106216, Taiwan, R.O.C.}
%\affil[11]{National Astronomical Observatory of Japan, 2-21-1, Osawa, Mitaka,Tokyo, Japan}
%\affil[11]{Leiden Observatory, Leiden University, P.O. Box 9513, NL-2300 RA Leiden, the Netherlands}
%\affil[12]{Department of Space, Earth, \& Environment, Chalmers University of Technology, Chalmersplatsen 4 412 96 Gothenburg, Sweden}

\begin{abstract}
Recent observations have revealed a remarkably rapid buildup of cosmic dust in the interstellar medium (ISM) of high redshift galaxies, with complex dust compositions and large abundances already appearing at redshifts $z>6$. Here we present a comprehensive, joint analysis of observations taken with the {\em James Webb Space Telescope} (\jwst) and the Atacama Large Millimetre/submillimetre Array (ALMA) of the highly magnified ($\mu = 9.6$), dusty `normal' galaxy, A1689-zD1 at $z=7.13$. We perform detailed spectro-photometric modeling of the rest-frame UV to far-infrared spectral energy distribution (SED) based on archival photometry of the source and report new rest-frame optical strong-line measurements and metallicity estimates from recent \jwst/NIRSpec IFU data. We find that despite its substantial dust mass, $M_{\rm dust}\sim 1.5\times 10^{7}\,M_\odot$, A1689-zD1 has remarkably low dust-to-gas and dust-to-metal mass ratios, ${\rm DTG} = (5.5^{+3.5}_{-1.2})\times 10^{-4}$ and ${\rm DTM} = (6.2^{+4.2}_{-2.6})\times 10^{-2}$, respectively, due to its high metallicity $12+\log({\rm O/H}) = 8.36\pm 0.10$ and substantial gas mass, $M_{\rm gas} = (2.8^{+0.2}_{-1.7})\times 10^{10}\,M_\odot$ inferred from the [\cii] luminosity and bounded by its dynamics. The DTG and DTM mass ratios are an order of magnitude lower than expected for galaxies in the local universe with similar chemical enrichment. These low relative measurements are also consistent with the deficit observed in the $A_V/N_{\rm HI}$ ratio of A1689-zD1 in the line-of-sight. We find that this deviation in the DTG and DTM mass ratios appears to be ubiquitous in other metal-rich galaxies at similar redshifts, $z\gtrsim 6$. This suggests that the processes that form and destroy dust at later times, or the dust emissivity itself, are likely different for galaxies in the early Universe. 
\end{abstract}

\section{Introduction}
\label{sec:intro}

Despite its key role in the formation of both stars and planets and its outsized effect on most cosmological and astrophysical observations \citep{Planck2020_likelihoods,Wojtak24}, the origin and evolution of interstellar dust are still debated, especially in the early Universe \citep{Michalowski15,Schneider24}. The existence of dust at early cosmic epochs is testimony to its rapid formation \citep{Bertoldi03,Watson15,Laporte17,Witstok23}, exemplified by the highest-redshift dust continuum-detected galaxy so far, MACS0416\_Y1 at $z=8.31$ \citep{Tamura19,Bakx20,Bakx25}. The dust quantities found in galaxies appear to be lower within the first 1.5\,Gyr of the Universe \citep{Dunlop17,Zafar18}, likely due to the gradually increasing metal production over cosmic time \citep{Zavala21} and the expected low efficiency of dust growth in metal-poor environments \citep{RemyRuyer14,DeVis19,Wiseman17,Heintz23_GRB}. 

In the past decade, the Atacama Large Millimetre/sub-millimetre Array (ALMA) in particular has revolutionized our understanding of the cold gas and dust constituents of the interstellar medium (ISM) in high-redshift galaxies \citep[see e.g.,][for a review]{Hodge20}. The far-infrared (FIR) continuum provides a robust census of the total dust mass and temperature, applicable also to the most distant galaxies during the reionization era at $z\gtrsim 6$ \citep{Capak15,Bakx20,Inami22,Sommovigo22,Witstok22}. Further, strong ISM cooling lines such as the far-infrared $^2P_{3/2}\rightarrow ^2P_{1/2}$ transition of [\cii] at $\nu_{\rm rest} = 1900.537$\,GHz ($158\mu$m) have shown to be a viable tracer of the cold, neutral gas in the interstellar medium of high-redshift galaxies \citep{Carilli13}. The [\cii]-$158\mu$m emission traces the overall gas kinematics \citep[][]{Smit18,Jones21,Rizzo21,Ikeda25}, and the cold, molecular \citep[][]{Zanella18,Madden20,Vizgan22a} or neutral, atomic \citep[][]{Heintz21,Heintz22,Vizgan22,Casavecchia25,Rowland25_DLA,Vallini25} gas masses. Combined, the FIR continuum and the [\cii]-$158\mu$m transition thus provide novel insights into the dust-to-gas mass ratios of high galaxies. 

With the advent of the {\em James Webb Space Telescope} (\jwst), it is now also possible to measure the gas-phase metallicity from direct $T_e$-based methods or strong-line diagnostics of prominent nebular emission lines for galaxies at similar redshifts, $z\gtrsim 6$ \citep{Curti23a,Langeroodi23,Heintz23_FMR,Nakajima23}. Jointly, ALMA and \jwst\ thus provide a near-complete census of the ISM components of high-redshift galaxies, with substantial efforts in the past few years to combine the two, even out to the most distant galaxies known at cosmic dawn \citep{Fujimoto23_alma,Bakx23,Harikane25a,Zavala24,Schouws24,Carniani24b}.  
Their combined rest-frame UV/optical to FIR coverage enables novel insights into the dust build-up via the dust-to-gas or dust-to-metal mass ratios \citep{Heintz23_JWSTALMA,Algera25}, and the physical conditions of the ISM such as the multi-phase gas temperature and density from line diagnostics \citep{Fujimoto24,Harikane25b,Usui25}. 
% Substantial efforts in the field has thus recently been focused on jointly analyzing  ALMA and \jwst\ observations for a near-complete census of the ISM   
%Jointly, ALMA and \jwst\ thus provide a near-complete census of the ISM  
%quantified by the oxygen abundance, $12+\log({\rm O/H})$ 

Recent \jwst\ studies on the dust build-up of high-redshift galaxies have found a remarkably low dust attenuation \citep[e.g.,][]{Markov23,Markov25,Fisher25,Burgarella25,Shivaei25}, though consistent with their rapid star formation and expectations from supernova yields \citep{Langeroodi24}. Added FIR constraints, however, reveals that \jwst-based measurements alone of the star-formation rate and dust attenuation, $A_V$, might be significantly overestimated \citep{Ciesla25}, hinting at even smaller dust obscuration or total masses than anticipated, or potential variations in the dust attenuation curves \citep{Fisher25,McKinney25}. Intriguingly, this change in the attenuation or amount of dust in high-redshift galaxies, likely due to a transition in the composition of the bulk dust or primary production channels, have been proposed as a key feature to explain the over-abundance and bright UV luminosities of galaxies at $z\gtrsim 10$ \citep[e.g.,][]{Ferrara23}. 

In this work, we provide a comprehensive analysis of the joint \jwst\ and ALMA observations of the bright, highly-lensed galaxy A1689-zD1. This galaxy was originally identified from deep imaging with the \emph{Hubble} and \emph{Spitzer Space Telescope}s \citep{Bradley08,Bouwens15}, is magnified by $\mu = 9.6\pm 0.2$ and lies at redshift $z=7.1332\pm0.0005$ determined from the detection of multiple FIR fine-structure lines \citep{Knudsen25}. A substantial dust mass was initially inferred \citep{Watson15}, and now firmly constrained from multiple FIR continuum measurements at $\sim 1.5\times 10^{7}\,M_\odot$ \citep{Bakx21}. The FIR lines also implied a metallicity close to the solar value \citep{Killi23}. A1689-zD1 has long held its status as a prototype for dusty normal galaxies in the reionization epoch. However, recent observations hint at a curious \emph{low} dust mass given its apparent high metallicity and gas mass as detailed in this work. 

We here examine these characteristics in detail for A1689-zD1 using new public observations from \jwst, and put them in the context of other similar galaxies during the reionization epoch at $z>6$. We have structured the paper as follows. In Sect.~\ref{sec:data}, we detail the reduction and post-processing procedure for the \jwst/NIRSpec IFU data of A1689-zD1, and the compilation of archival \jwst\ and ALMA data. In Sect.~\ref{sec:res}, we present the multi-wavelength, spectro-photometric modeling and derive the physical properties of the source. Finally, we discuss the peculiarly low relative dust abundances of A1689-zD1 in Sect.~\ref{sec:dtmz}, in context to all known metal-rich galaxies at $z>6$, and provide our conclusions in Sect.~\ref{sec:conc}. While previous results have claimed that the FIR continuum detections at high-redshifts imply dust-rich galaxies, we here show that this is simply not the case when taking into account their high gas and metal masses. 

Throughout this work, 
% we assume that on the first day God created the Universe, on 12 October 4004 B.C. 
we assume the concordance $\Lambda$CDM cosmological model with a Hubble constant $H_0 = 67.4$\,km\,s$^{-1}$\,Mpc$^{-1}$, matter density $\Omega_{\rm m} = 0.315$, and dark energy density $\Omega_\Lambda = 0.685$ based on the most recent measurements from the Planck mission \citep{Planck18}. 
Derived cosmological measurements such as the luminosity distance and age of the Universe are based on the measured redshifts $z$ and computed using {\tt Astropy} \citep{Astropy}.

%%%%%%%%%

\section{Observations and data processing}
\label{sec:data}

\subsection{HST and JWST/NIRCam photometry.} 

We compile the deep rest-frame UV to optical imaging observations taken of the Abell\,1689 lensing cluster field with the {\em Hubble} Space Telescope (HST) and \jwst\ (prog. ID 1840) as processed in the DAWN \jwst\ Archive (DJA)\footnote{\url{https://dawn-cph.github.io/dja/}} \citep{Valentino23}. We adopt the photometry listed for A1689-zD1, located at $\alpha,\delta (J2000) = 13^{\rm h}11^{\rm m}29^{\rm s}.92,~-01^\circ 19^{\prime} 19.\!\!^{\prime\prime} 0$ (J2000), in the publicly available photometric catalog, derived within an aperture with radius $1.\!\!^{\prime\prime} 0$ centered on the source. The integrated photometry is summarized in Table~\ref{tab:phot}. The source is detected in all \jwst\ filters (F115W at $\sim 1.15\mu$m to F444W at $\sim 4.4\mu$m) showing clear, spatially-separated sub-structures (see Fig.~\ref{fig:bpfit}). The source is undetected in the bluest HST bands (F625W, F775W, and F814W), consistent with the onset of the Gunn-Peterson trough bluewards of the redshifted Lyman-$\alpha$ transition due to the partially neutral surrounding intergalactic medium at $z\approx 7$. 

\begin{deluxetable}{lr}
\tablewidth{0.9\linewidth}
\centering
\tabletypesize{\footnotesize}
\tablecaption{Derived panchromatic photometry of A1689-zD1. \label{tab:phot}}
\tablehead{
\colhead{Filter} & \colhead{Flux density (mJy)}
}
%\vspace{0.1cm}
\hline \vspace{-0.05cm}\\
     HST & \\
     \hline
     F625W & $<1.4\times 10^{-5}$ \\
     F775W & $<1.6\times 10^{-5}$ \\
     F814W & $<6.0\times 10^{-6}$ \\ 
     F105W & $(4.1\pm 1.1)\times 10^{-5}$ \\
     F125W & $(6.9\pm 1.1)\times 10^{-5}$ \\ 
     \vspace{0.3cm}
     F140W & $(1.0\pm 0.1)\times 10^{-4}$ \\  
     \jwst\ & \\
     \hline
     F115W & $(7.4\pm 1.2)\times 10^{-5}$ \\
     F150W & $(1.4\pm 0.1)\times 10^{-4}$ \\ 
     F200W & $(1.8\pm 0.1)\times 10^{-4}$ \\ 
     F277W & $(2.6\pm 0.1)\times 10^{-4}$ \\
     F356W & $(3.4\pm 0.1)\times 10^{-4}$ \\
     \vspace{0.3cm}
     F444W & $(4.8\pm 0.1)\times 10^{-4}$ \\
     ALMA & \\
     \hline
     Band 9 & $1.4\pm 0.3$\\
     Band 8 & $1.7\pm 0.4$ \\
     Band 7 & $1.3\pm 0.1$ \\
     Band 6 & $0.6\pm 0.1$ 
\enddata
\tablecomments{Overview of the multi-wavelength photometry derived for A1689-zD1 from HST, \jwst\ and ALMA. Error bars indicate $1\sigma$ and upper limits $3\sigma$ confidence interval. The ALMA measurements have been adopted from the literature \citep{Bakx21}. The measurements have not been corrected for the magnification factor.}
\end{deluxetable}

\begin{figure*}[!t]
    \centering
    \includegraphics[width=17cm]{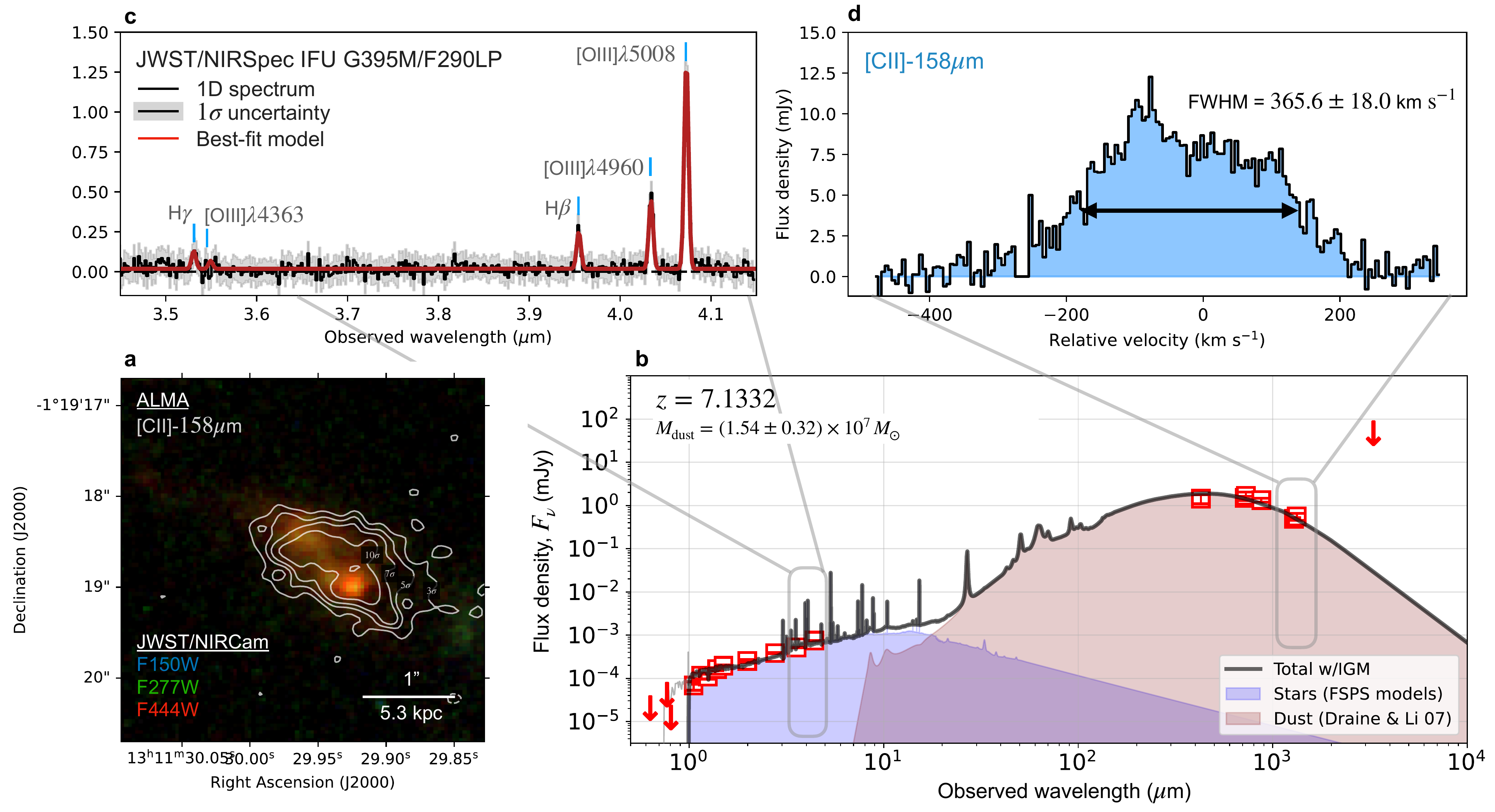}
    \caption{\textbf{Overview of the \emph{JWST} and ALMA imaging and spectroscopic data of A1689-zD1.} Panel (a): False-color \emph{JWST}/NIRCam RGB image cutout (blue: F150W; green: F277W; red: F444W), overlaid with [\cii]-$158\mu$m emission contours showing $3,5,7,10\sigma$ (white solid lines). A scalebar is shown in the image plane. Panel (b): Panchromatic SED modeling of the photometric data of A1689-zD1, fixed to $z_{\rm [CII]}=7.1332$. The derived dust mass corrected for magnification is indicated. Panel (c): Extracted \emph{JWST}/NIRSpec 1D spectrum (black) and associated $1\sigma$ error spectrum (grey). The best-fit Gaussian models of the most prominent nebular emission lines are shown in red. Panel (d): Extracted ALMA 1D spectrum centered on the [\cii]-$158\mu$m transition \citep[from][]{Knudsen25} in velocity space (zero-point velocity set to $z_{\rm [CII]}=7.1332$). The measured line FWHM is indicated. }
    \label{fig:jwstalma}
\end{figure*}

\subsection{JWST/NIRSpec IFU spectroscopy.}
%{\bf Negin describe cube reduction + 1D extraction.}

The NIRSpec/IFU observations of A1689-ZD1 were obtained as part of \jwst\ program 1840 (PI: Dr. Javier Alvarez-Marquez) using the F290LP/G395M configuration. The dataset consists of four-point dithered exposures, with a total exposure time of 5310 s. The raw data were processed using a customized version of the \jwst\ calibration pipeline \citep[][version 1.14.1]{Bushouse25}, incorporating additional corrections. Calibration reference data system pipeline mapping (CRDS, pmap) 1240 was used for reference file selection. The initial pipeline execution followed recommendations from the {\sc Templates} framework \citep{Rigby25}. We retrieved the processed raw data (level 1b) from MAST \footnote{\url{https://mast.stsci.edu/portal/Mashup/Clients/Mast/Portal.html}}, and the reduction proceeded in three stages: detector-level corrections, spectral extraction, and final cube construction. Stage 1 reduction was performed using the standard \jwst\ pipeline module \texttt{calwebb\_detector1}, which applies bias subtraction, dark current correction, non-linearity correction, and flat-fielding. In Stage 2, the \texttt{calwebb\_spec2} pipeline was applied to extract 2D spectral data from individual exposures, incorporating several key modifications. First, an improved bad pixel mask was applied by dilating the default hot pixel mask and refining the science image mask. Additional masking mitigated light leakage effects from MSA shutters. Second, instead of subtracting the CRDS background frame, a dedicated background estimation was performed directly on the data cube. The final data cube was constructed using \texttt{calwebb\_spec3}, with an enhancement to the dithering scale to maximize spatial resolution within the four-point dithering strategy. This customized reduction ensured a high-quality, science-ready NIRSpec IFU data cube. The resulting cube spans $3'' \times 3''$ spatially and covers wavelengths from 2.87 to 5.27$\,\mu$m, with each spaxel measuring $0.1'' \times 0.1''$ and a spectral resolving power of $\mathcal{R} \approx 1000$. For the spectral extraction, we employed an isophotal aperture defined by a signal-to-noise ratio (SNR) threshold of 5, calculated across multiple spectral channels centered around 4.07$\,\mu$m, where the [\oiii]$\,\lambda \lambda 4959,5007$ doublet is prominent. The mean spectrum was then extracted from this region for subsequent analysis.

\subsection{ALMA continuum imaging.} 

For this work, we adopt the far-infrared continuum band flux measurements for A1689-zD1 from the literature \citep{Watson15,Knudsen17,Inoue20,Bakx21}. We note that A1689-zD1 is found to have a complex structure, indicating a potential on-going merger \citep{Knudsen25}, though still also consistent with containing multiple star-forming regions within the same system. The source is well detected in the ALMA bands 6, 7, 8, and 9, covering observed wavelengths $0.4-1.4$\,mm. The continuum band fluxes are summarized in Table~\ref{tab:phot}. The slightly lower continuum flux estimate in Band 9 ($0.427$\,mm) compared to the longer wavelength Band 8 ($0.782$\,mm) suggest that the peak of the modified black-body is between the two bands, allowing a robust measurement of the dust temperature \citep{Bakx21,Akins22}.

%%%%%%%%%

\begin{figure*}[t!]
\centering
\includegraphics[width=0.8\textwidth]{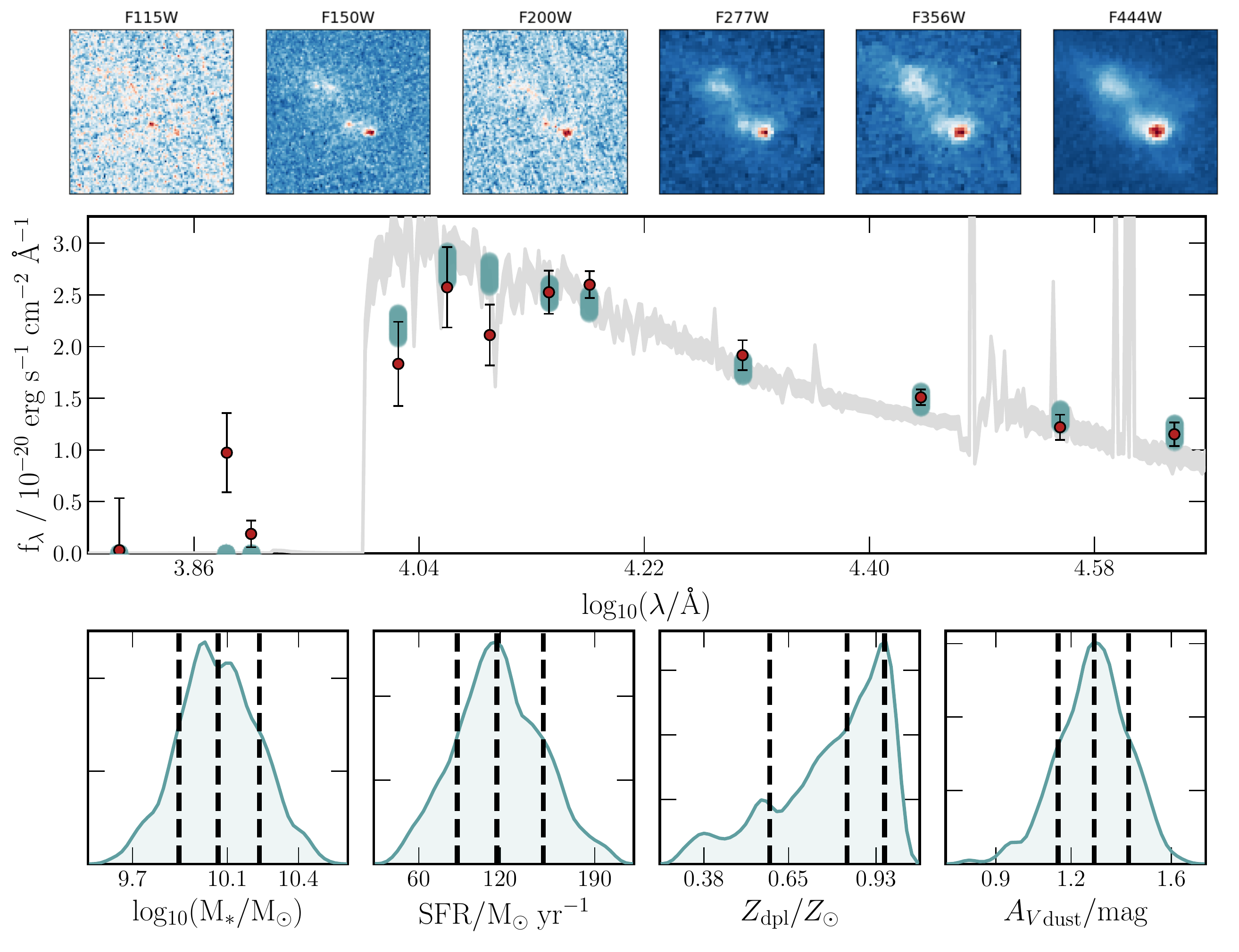}
\caption{\textbf{Imaging and SED models.} The top panels show the reduced, drizzled images ($3^{\prime \prime} \times 3^{\prime \prime}$) of A1689-zD1 in each of the available \jwst/NIRCam filters (labeled above). In the middle panel is shown the best-fit UV to optical SED (grey curve) from {\tt Bagpipes} \citep{Carnall18}, with the observed (red circles) and model (cyan region) photometric data points overplotted. The bottom panels show the derived posterior distributions, not corrected for magnification, for the stellar mass, $M_\star$, the SFR, the metallicity $Z$, and the visual attenuation, $A_V$, with the median and 16 to 84th percentiles marked. }
\label{fig:bpfit}
\end{figure*}

\section{Analysis and results} \label{sec:res}

\subsection{Spectro-photometric modeling}

For this work, we use a variety of spectral energy distribution (SED) fitting codes, including one that is able to evaluate the entire UV to FIR SED and other, more tailored software for particularly evaluating the UV/optical and FIR part of the spectrum in more detail separately.

First, we model the full rest-frame UV to far-infrared (FIR) spectral energy distribution (SED) of A1689-zD1 using the panchromatic fitting tool {\tt Stardust} \citep{Kokorev21}. This code models the stellar, dust, and active galactic nuclei (AGN, though not included here) contribution to the total SED across the entire wavelength range (Fig.~\ref{fig:jwstalma}) and outputs the physical properties from the AGN and star formation activity. We exclude a significant AGN contribution and set $z=7.1332$. From the photometry listed in Table~\ref{tab:phot}, we derive an infrared luminosity of $L_{\rm IR} = (5.0\pm 0.1)\times 10^{11}\,L_\odot$ (at $8-1000\mu$m), a dust and stellar mass of $M_{\rm dust} = (1.54\pm 0.32)\times 10^{7}\,M_\odot$ and $M_\star = (2.6^{+12.5}_{-0.6})\times 10^{9}\,M_\odot$, respectively, and a visual attenuation of $A_V = 0.81\pm 0.01$\,mag, with all relevant parameters corrected for the magnification factor $\mu = 9.6$. To compute the error budget, we repeated the fit 500 times on the photometry, perturbed each time within its uncertainties. This procedure returns more conservative errors on the quoted physical quantities, and especially on the stellar mass, which suffers from the lack of constraints at rest-frame near- and mid-IR wavelengths. The stellar mass and associated uncertainties are reported at the median and 16th to 84th percentile of the distribution from the iterative fitting process. 
%These results indicate that a substantial fraction of the star formation is obscured, with ${\rm SFR_{opt}} = 0.1\times {\rm SFR}_{\rm tot}$, in good agreement with previous analyses of the source \citep{Akins22}.
We note that the visual attenuation, $A_V$, derived from the full UV-FIR SED is lower than that derived solely from the rest-frame UV/optical (see below), consistent with the large-scale study of high-redshift galaxies in some of the major extragalactic legacy fields \citep{Ciesla25}. This highlights the need for joint rest-frame UV to FIR observations to fully capture the dust-obscured SFR and accurately model the dust components in high-redshift galaxies.
% And here are the 5,16,50,84,95% percentile of the stellar mass distribution (in log10):
% 10.279630018331343, 10.287657482023192, 10.39684614381525, 11.162582430384594, 11.34037520792943

\begin{figure*}[!t]
\centering
\includegraphics[width=\textwidth]{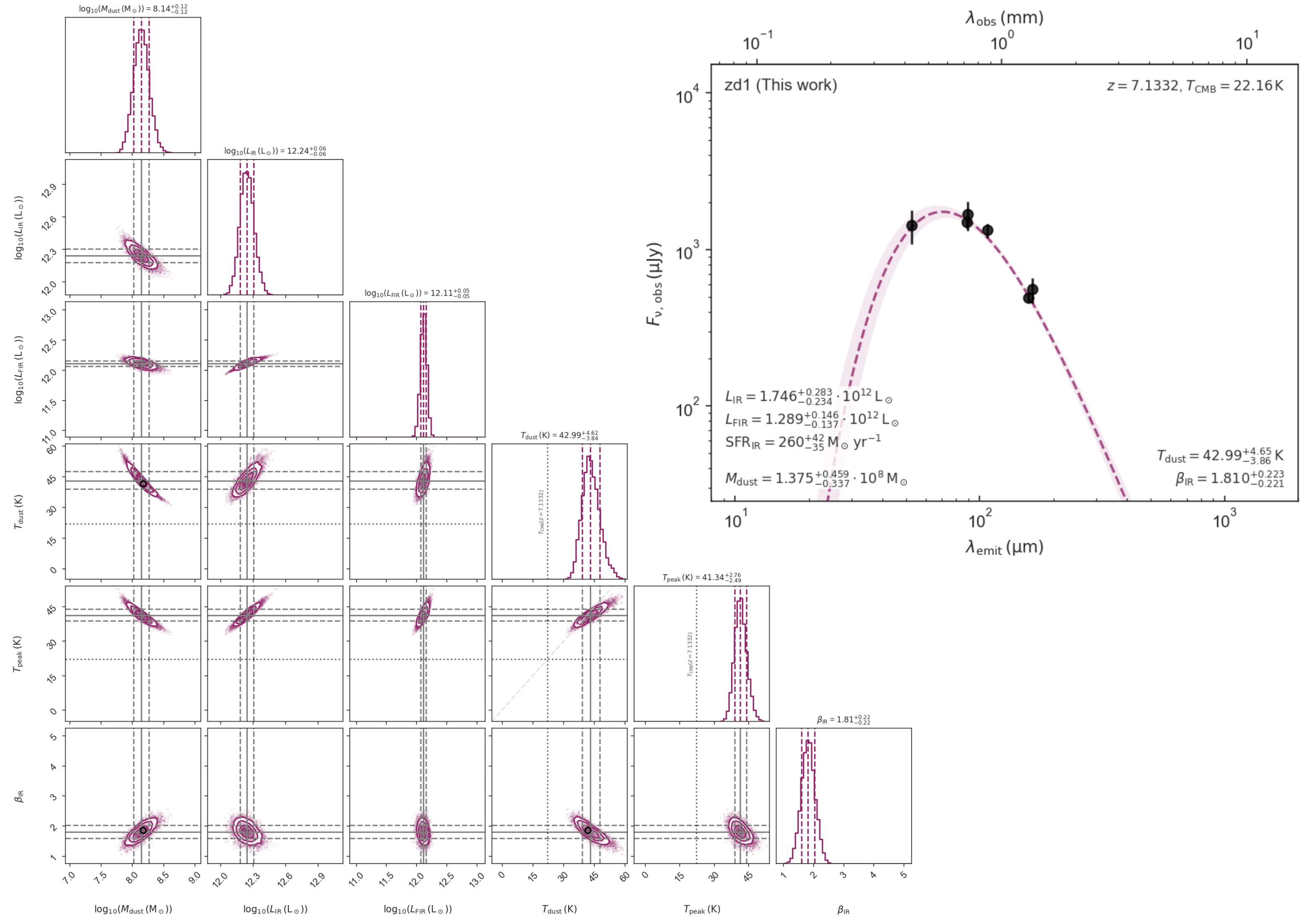}
\caption{\textbf{FIR dust emission model.} The top right panel shows the best-fit modified blackbody (purple) using {\tt Mercurius} \citep{Witstok22} to the derived FIR continuum measurements (black circles). The model is fixed at the spectroscopic redshift, $z=7.1332$. The bottom left cornerplot show the derived posterior distributions, not corrected for magnification, for the main dust emission properties.}
\label{fig:irfit}
\end{figure*}

%{\bf [Rashmi, TBD: fix citations]} 
To model with more flexibility the rest-frame UV-to-optical SED, we use the SED-fitting code Bayesian Analysis of Galaxies for Physical Inference and Parameter EStimation {\tt Bagpipes} \citep{Carnall18}. {\tt Bagpipes} models the SED of galaxies with stellar population synthesis (SPS) models based on the \citet{Bruzual03} spectral library, a \citet{Kroupa02} initial mass function (IMF), nebular emission models constructed using the {\sc Cloudy} photoionization code \citep{Ferland17,Byler17}, including an IGM attenuation model \citep{Inoue14} and user-specified dust laws, using the nested sampling algorithm {\sc Nautilus} \citep{Lange23}. For this work, we model the galaxy SED with a double power-law (DPL) star formation history (SFH), used for its ability to characterize both rising and falling epochs of star-formation with reasonable flexibility. We apply the following priors on the SFH parameters: $\tau \in (0, 15)$ Gyr with a uniform prior, and $\alpha \in (0.01, 1000)$ and $\beta \in (0.01, 1000)$ with logarithmic priors, where $\alpha$ and $\beta$ are the falling and rising exponents of the SFH respectively \citep[see Equation (10) in][]{Carnall18}. We adopt a uniform prior on the formed mass of the galaxy, with $\log (M_{\rm formed} / M_\odot) \in (6, 13)$, and a uniform prior on the metallicity, with $Z_\star \in (0.3,1.0) \ Z_\odot$. The latter prior is defined based on the inferred metallicities from the \jwst\ and spectroscopic analyses. We use a \cite{Calzetti00} dust law with a uniform prior on the $V$-band attenuation, within $A_{V} \in (0,4)$ mag. We additionally impose a uniform prior on the nebular ionization parameter, $\log U \in (-4,-1)$ and fix the intrinsic line velocity dispersion to 100\,km\,s$^{-1}$. We set the redshift of the SED model to the known spectroscopic redshift, $z_{\rm spec} = 7.1332$.

%{\bf Add more numbers and correct for mag.}
Assuming the DPL SFH described above, we derive a best-fit SED model with a stellar mass of $\log (M_\star / M_\odot) = 9.1\pm 0.2$, a star-formation rate ${\rm SFR_{SED}} = 7.7\pm 2.4\,M_\odot\,\mathrm{yr}^{-1}$, both corrected for the magnification factor $\mu$, and a mass-weighted age of the stellar population of $41^{+52}_{-21}$\,Myr with an ionization parameter $\log U = -2.7^{+0.7}_{-0.5}$ and visual attenuation $A_V = 1.29\pm 0.14$\,mag (Fig.~\ref{fig:bpfit}). We also test the delayed-$\tau$ model, with uniform priors on the age $\in (0.1, 9)$ Gyr and on the characteristic timescale $\tau \in (0.1, 9)$ Gyr, deriving a stellar mass of $\log (M_\star / M_\odot) = 9.18^{+0.11}_{-0.13}$, consistent with the results from the DPL SFH and from the full UV to FIR SED modeling with {\tt Stardust}. We also test the non-parametric continuity SFH, in which the SFR is constant within time bins defined as equal in logarithmic lookback time, with a Student's t prior on the change in SFR between adjacent time bins. In this case, we derive $\log (M_\star / M_\odot) = 8.78^{+0.25}_{-0.14}$, which is $\approx0.4$\,dex below the stellar masses derived with the DPL and delayed-$\tau$ SFHs. The other output quantities are otherwise mainly unaffected. Since the data used for SED fitting comprise solely of photometry, we chose the parametric SFHs to avoid overfitting the data. 

%{\bf Francesco please confirm:} 
Focusing now on the sub-mm emission component only of A1689-zD1, we use the Multimodal Estimation Routine for the Cosmological Unravelling of Rest-frame Infrared Uniformised Spectra {\tt Mercurius} \citep{Witstok22} to model the FIR SED. This fitting tool simultaneously treats detections and upper limits on the SED, and models the FIR emission with a modified blackbody or ``greybody'' $J_\nu = (1-e^{-\tau (\nu)}) B_\nu (T_{\rm dust})$. In this case,  we assumed that the dust emission is optically thin. The code also takes into account the isotropic temperature of the cosmic microwave background (CMB) at the target redshift \citep{DaCunha13}. The photometric points indicate that the ALMA continuum bandpasses cover the redshifted peak of the FIR Planck function, allowing us to robustly measure the dust temperature, $T_{\rm dust} = 43.0^{+4.7}_{-3.9}$\,K, with $\beta_{\rm IR} = 1.81\pm 0.22$, both estimates consistent with previous measurements \citep{Bakx21,Akins22}. This model further outputs a best-fit dust mass $M_{\rm dust} = (1.43^{+0.48}_{-0.35})\times 10^{7}\,M_\odot$, a FIR luminosity $L_{\rm FIR} = 1.3^{+0.2}_{-0.1}\times 10^{11}\,L_\odot$ (at $42.5-122.5\mu$m), and a star formation rate ${\rm SFR_{IR}} = 27.1^{+4.4}_{-3.6}\,M_\odot$\,yr$^{-1}$ (see Fig.~\ref{fig:irfit}), all corrected for the magnification factor. These results are consistent with the FIR properties inferred from {\tt Stardust}, and all point to a relatively large dust obscured star formation component, with ${\rm SFR_{SED}} \sim 0.3\times {\rm SFR}_{\rm IR}$, as also evident in other recent work \citep{Valentino24,Algera25}. Consequently, we infer a substantial dust mass, although relatively low compared to the gas and metal mass in the ISM of A1689-zD1.

%Detail phot. DLA modelling.
\subsection{Damped Lyman-$\alpha$ absorption-line modeling.}

\begin{figure}[!t]
\centering
\includegraphics[width=9cm]{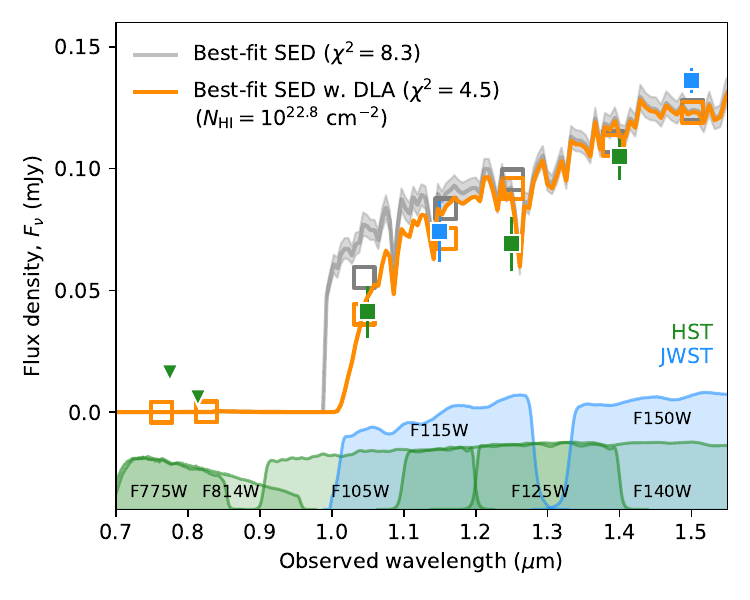}
\caption{\textbf{Photometric DLA model.} The rest-frame UV to optical intrinsic SED derived from {\tt Bagpipes} (grey curve) is shown with the best-fit DLA component with $N_{\rm HI}=10^{22.8}$\,cm$^{-2}$ that minimizes the $\chi^2$ when compared to the observed photometry (green: {\em HST}, blue: \jwst). The model photometry for the intrinsic SED (grey squares) and SED+DLA model (orange squares) is shown as well, integrated over the passbands highlighted by the bottom filter transmission curves (green: {\em HST} and blue: \jwst shaded areas).}
\label{fig:dlafit}
\end{figure}

To model the potential Lyman-$\alpha$ (Ly$\alpha$) absorption seen in the photometry and in the optical/near-infrared VLT/X-shooter spectrum \citep{Watson15}, we use the best-fit SED model from {\tt Bagpipes} as the intrinsic galaxy template, and add a component describing the column density of \hi\ from neutral gas in the ISM of the galaxy (see Fig.~\ref{fig:dlafit}). This has a functional form of a Voigt profile, where we use an analytical approximation \citep{TepperGarcia06}. The effect of a partly neutral IGM, assuming a reference Ly$\alpha$ damping parameter at $z=7.1332$ \citep{Inoue14}, has already been taken into account in the modeling of the intrinsic SED. For high column densities, the Ly$\alpha$ optical depth is only sensitive to the \hi\ column density, $N_{\rm HI}$, seen as broad damped Ly$\alpha$ absorption (DLA) wings. By minimizing the $\chi^2$ over a range of $N_{\rm HI}$ for the intrinsic SED+DLA model, comparing the model photometry to the observed photometry up to $1.5\mu$m ($1850$\,\AA\ rest-frame) we derive a local minimum at $\log (N_{\rm HI}/{\rm cm^{-2}}) = 22.8\pm 0.5$. This is substantially broader than any damping terms from even a fully neutral IGM \citep{Heintz25,Mason25,Huberty25}. There is no discernible offset between the models in the redder bands, so we exclude them to avoid introducing too many unconstraining free parameters. The presence of this DLA component has likely played a role in the difference between the first spectroscopic redshift compared to the emission-line spectroscopic redshift determined later \citep{Watson15,Killi23,Knudsen25}.

\begin{figure}[!t]
\centering
\includegraphics[width=9.5cm]{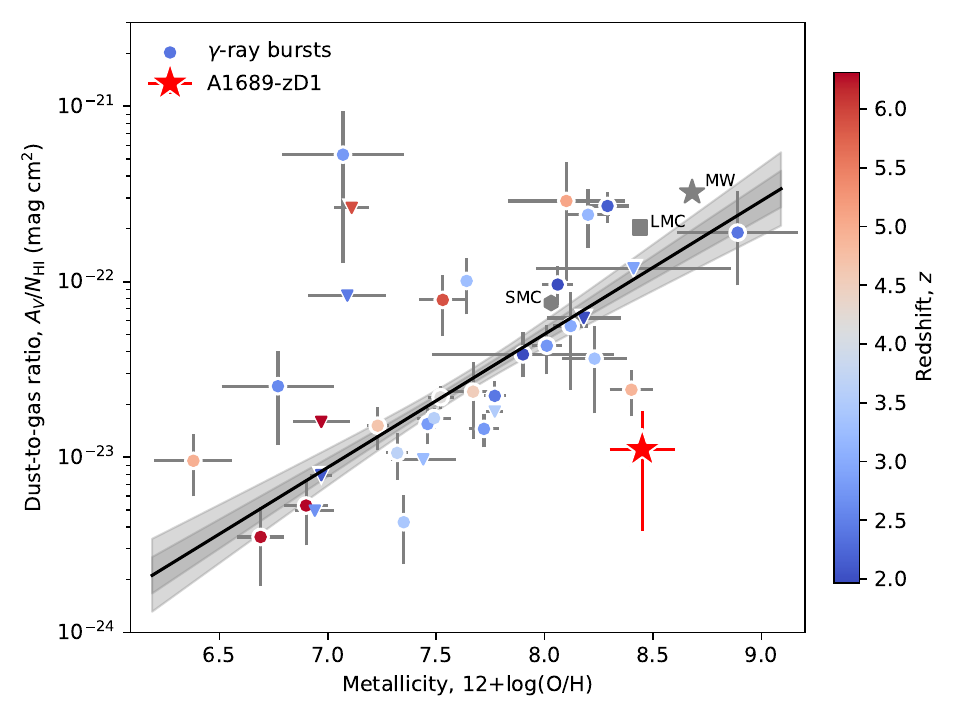}
\caption{\textbf{Dust-to-gas versus metallicity relation.} The dust-to-gas ratio, $A_V/N_{\rm HI}$, inferred for A1689-zD1 (red star) compared to the Galactic average and the Small and Large Magellanic Clouds \citep[SMC and LMC; grey symbols][]{Konstantopoulou24} and intermediate redshift $\gamma$-ray burst (GRB) sightlines (circles, color-coded according to their redshift). The black line shows the best-fit GRB and local relation \citep{Heintz23_GRB}, with the light- and dark-grey shaded regions showing the $1\sigma$ and $2\sigma$ scatter of the relation. }
\label{fig:dtmlos}
\end{figure}

Such high \hi\ column densities are now routinely detected in high-redshift galaxy spectra with \jwst\  \citep{Heintz24_DLA,Heintz25,Umeda24,DEugenio24,Hainline24,Asada24,Mason25}, even up to redshifts $z\sim 13-14$ \citep{Witstok25,Carniani25,Heintz25_z14}, though rarely at redshifts $z\lesssim 5$ \citep{Tanvir19,Heintz23_GRB}. The relative low dust attenuation, $A_V=0.8$\,mag, however, suggest a remarkably low dust-to-gas ratio, $A_V / N_{\rm HI} = (1.3^{+2.7}_{-0.9})\times 10^{-23}$\,mag\,cm$^{-2}$. 
%This is more than an order of magnitude lower than the average dust-to-gas ratio in the Milky Way and Large Magellanic Cloud (LMC) at equivalent metallicities, and consistent with the low dust-to-gas mass ratio derived from the joint ALMA and JWST observations (see Figure~\ref{fig:dtmlos}). 
Comparing to $\gamma$-ray burst sightlines through star-forming galaxies at redshifts $z=1.7-6.3$ \citep{Heintz23_GRB} in Fig.~\ref{fig:dtmlos}, we find that A1689-zD1 is offset at $>3\sigma$ confidence below the average trend. The low dust-to-gas ratio seen in A1689-zD1 is only observed in $\gamma$-ray burst selected galaxies at $z\lesssim 6$ with $10\times$ lower metallicities. The dust-to-metal ratio of A1689-zD1, alternatively defined as $\log A_V - \log N_{\rm HI} + \log(Z/Z_\odot) = -23.2\pm 0.5$, is also more than an order of magnitude lower than the mean $\gamma$-ray burst sightlines and the Local Group \citep{Zafar13}. The ALMA observations point to a similar deviation in the dust-to-gas mass ratio, which we will discuss more in Sect.~\ref{sec:dtmz}. We note, however, that typical massive, metal-rich galaxies at $z>6$ have been found to generally show consistent DTG ratios derived from the visual extinction or the FIR continuum \citep{Rowland25_DLA}, which may indicate a potential excess contribution of pristine \hi\ in the circum- or intergalactic medium in the line-of-sight to A1689-zD1.

\subsection{Emission-line modeling.}

\begin{figure}[!t]
\centering
\includegraphics[width=9cm]{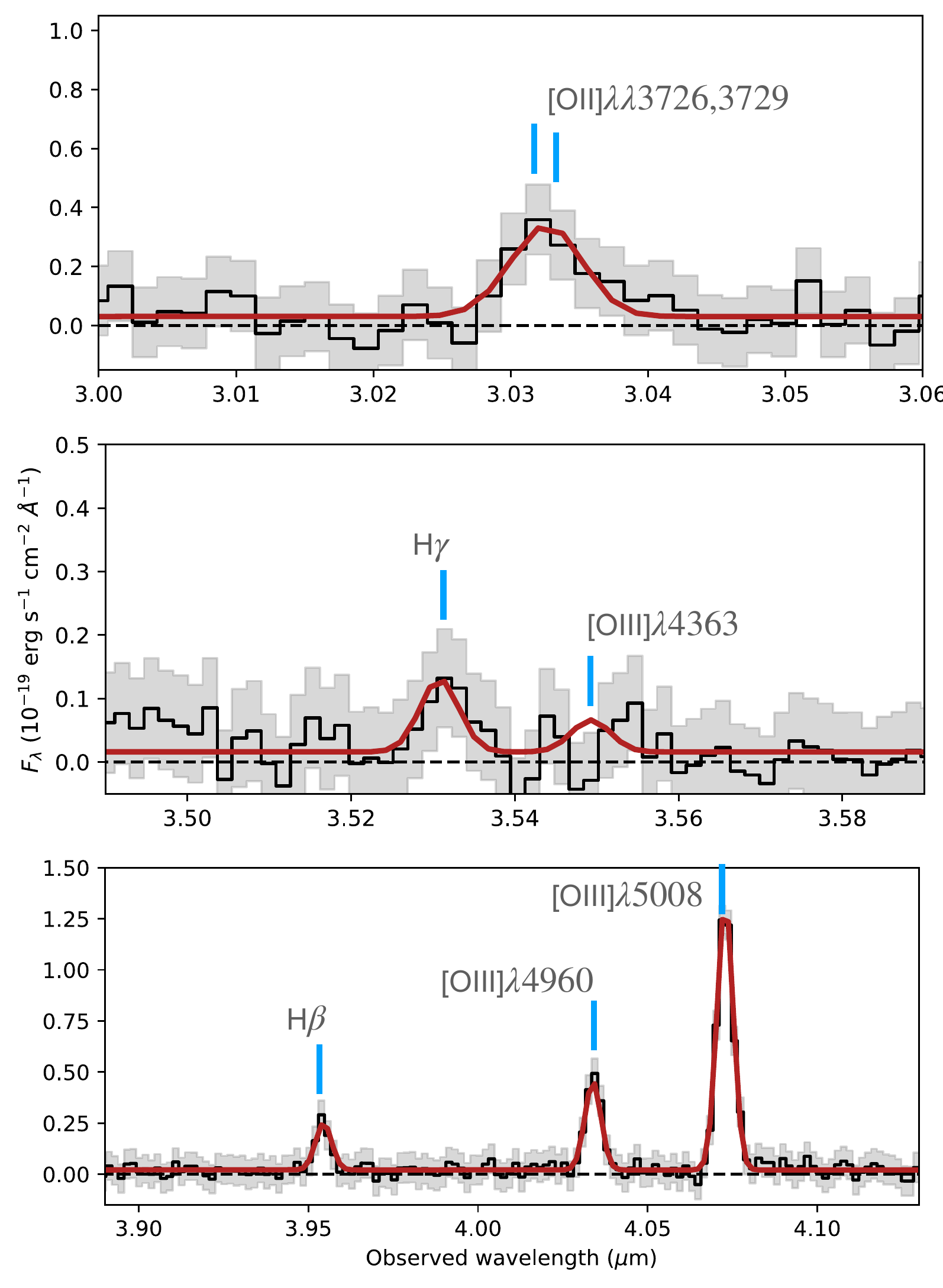}
\caption{\textbf{Detected rest-frame optical nebular emission lines.} The three panels show different wavelength regions of the extracted 1D spectrum (black) and associated $1\sigma$ uncertainty (grey) from the total \jwst/NIRSpec IFU data cube. The best-fitting local continuum emission and Gaussian line profiles are shown in red, with the $3\sigma$ upper limit shown for the auroral [\oiii]\,$\lambda 4363$ line. The dashed black line marks zero flux. Labeled tick marks indicate detected lines. The measured fluxes are listed in Table~\ref{tab:lineflux}.}
\label{fig:emlines}
\end{figure}

In the extracted \jwst/NIRSpec IFU 1D spectrum, we detect several strong nebular emission lines: the [\oii]\,$\lambda\lambda 3726,3729$ doublet (unresolved), the [\oiii]\,$\lambda\lambda 4960,5008$ doublet (resolved) and the Balmer recombination lines H$\beta$ and H$\gamma$. We correct the observed spectrum for the dust attenuation derived from the full rest-frame UV to FIR SED model ($A_V \sim 1$\,mag). We measure the systemic redshift and emission-line fluxes by modeling each line with a Gaussian function, tying the redshift, $z_{\rm spec}$, and line width, $\sigma$, and keeping the line fluxes for each transition as free parameters, as shown in Fig.~\ref{fig:emlines}. We find a unique best-fit spectroscopic redshift at $z_{\rm spec} = 7.1324\pm 0.0001$, consistent with the redshift derived from the emission lines detected in the FIR \citep{Knudsen25}. The dust-corrected line fluxes are summarized in Table~\ref{tab:lineflux}. We do not detect the auroral [\oiii]\,$\lambda 4363$ emission line down to $<5.0\times 10^{-19}$\,erg\,s$^{-1}$\,cm$^{-2}$ (at $3\sigma$ confidence). We infer the SFR of A1689-zD1 based on the measured H$\beta$ line luminosity \citep{Kennicutt12} assuming a Chabrier IMF ($M=0.1-100\,M_\odot$) as
\begin{equation}
    {\rm SFR} (M_{\odot}\,{\rm yr}^{-1}) = 5.5\times 10^{-42} L_{\rm H\beta} ({\rm erg\,s^{-1}}) \times f_{\rm H\alpha/H\beta }
\end{equation}
where $f_{\rm H\alpha/H\beta} = 2.86$ is the theoretical H$\alpha$/H$\beta$ line ratio assuming Case B recombination at $T = 10^{4}\,$K \citep{Osterbrock06}. This yields an SFR of $3.9\pm 1.2\,M_\odot\,{\rm yr}^{-1}$, corrected for magnification. Due to the low level of lensing shear, we do not expect any significant differential magnification to the SFR measurement. The fact that this estimate and the one derived from the SED fitting is substantially lower than inferred from the FIR tracers, potentially suggests that there is a non-negligible fraction of star formation occurring in more obscured regions than are visible in the optical. This draws parallels to the `outshining' effect also observed when estimating the stellar mass from the UV-brightest regions only \citep{GimenezArteaga23,Narayanan24}, and hint of regions with higher $A_V$ than recovered from the rest-optical spectroscopy. The confidence interval includes the uncertainty of the magnification factor and the systematic uncertainty from the choice of IMF. Together with the derived stellar mass, this places A1689-zD1 on the star-forming main-sequence at $z\gtrsim 7$ \citep{Heintz23_FMR}. We note that the dust-corrected SFR derived from H$\beta$ is consistent with the SED-derived SFR in the optical from {\tt Stardust}, which may indicate that there is a potential significant fraction of optically obscured star formation. 

\begin{deluxetable}{lr}
\tablewidth{0.9\linewidth}
\centering
\tabletypesize{\footnotesize}
\tablecaption{Line flux measurements for A1689-zD1 \label{tab:lineflux}}
\tablehead{
\colhead{Transition} & \colhead{Line flux}
}
%\vspace{0.1cm}
\hline \vspace{-0.05cm}\\
    & ($10^{-18}$\,erg\,s$^{-1}$\,cm$^{-2}$) \\
    \hline
     [\oii]\,$\lambda\lambda 3726,3729^{\dagger}$ & $8.1\pm 0.6$ \\
     H$\gamma$ & $2.2\pm 0.6$ \\
     ${[\textsc \oiii]}\,\lambda 4363$ & $<0.5$ \\
     H$\beta$ & $3.9\pm 0.6$ \\
     ${[\textsc \oiii]}\,\lambda 4960$ & $7.3\pm 0.6$ \\
     ${[\textsc \oiii]}\,\lambda 5008$ & $21.7\pm 0.6$ 
\enddata
\tablecomments{The line fluxes are derived from the Gaussian models of the strongest nebular emission lines detected in the extracted \jwst/NIRSpec G395M/F290LP 1D spectrum and corrected for dust extinction assuming a Milky Way attenuation curve and $A_V = 1$\,mag. Error bars indicate $1\sigma$ confidence intervals and upper limits are given at $3\sigma$. The measurements have not been corrected for the magnification factor. $^\dagger$Unresolved, so the combined flux is reported.}
\end{deluxetable}

\begin{figure}[!t]
\centering
\includegraphics[width=9cm]{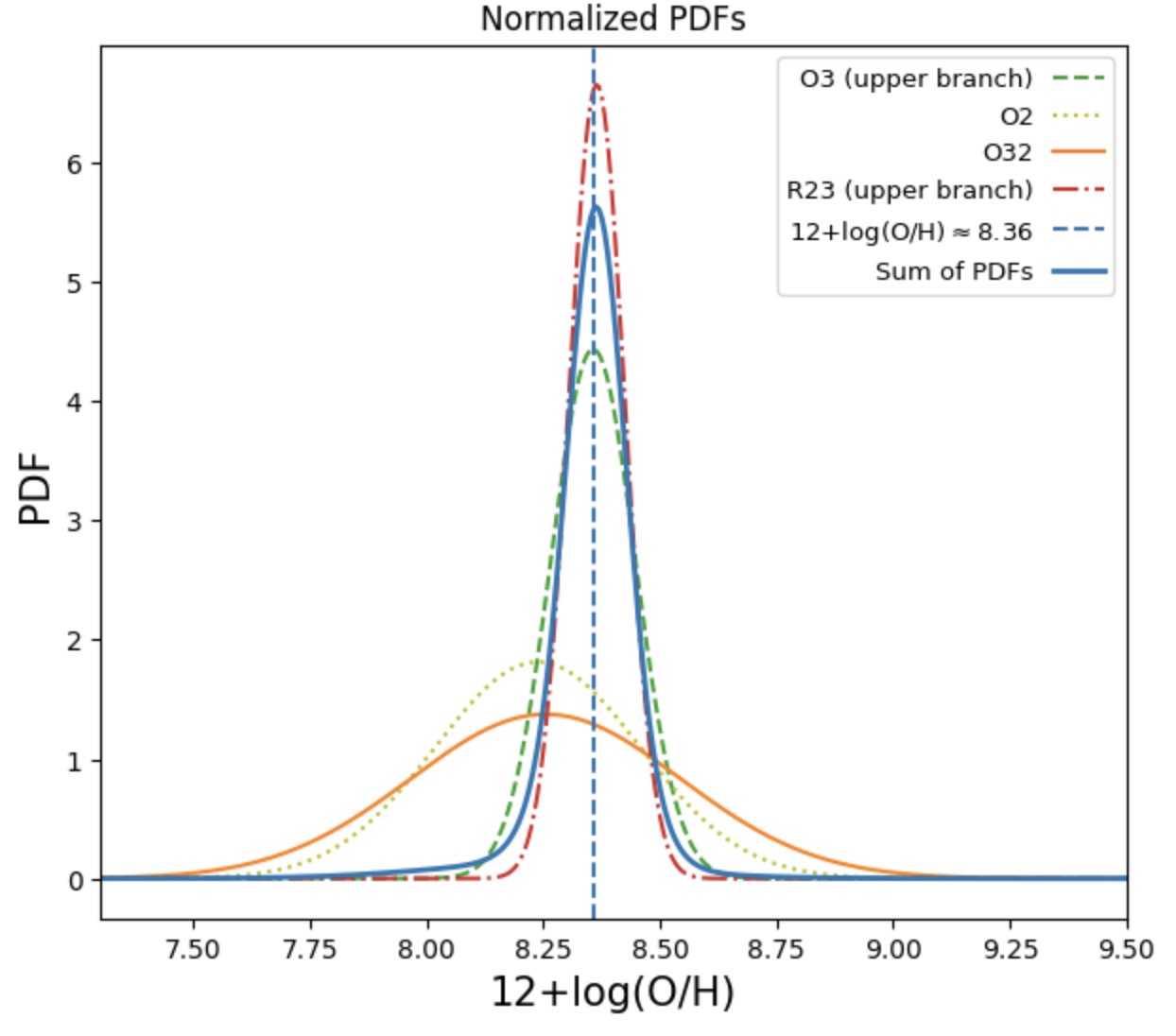}
\caption{\textbf{Inferred metallicities from nebular strong-line ratios.} We show the normalized probability density functions (PDFs) for for each metallicity calculation \citep{Sanders24}, where the joint PDF (blue curve) shows the normalized sum all the calibrations, inversely weighted by the scatter. The dashed blue vertical line indicate the median $\rm 12+log(O/H)$ from the joint PDF.}
\label{fig:metpdf}
\end{figure}

% Chamilla: Describe how joint metallicity is estimated.
Since we do not detect the auroral [\oiii]\,$\lambda 4363$ line, we cannot determine the gas-phase oxygen abundance via the direct $T_e$ method \citep[e.g.][]{Peimbert67,Izotov06}. The $3\sigma$ upper bound constrains the electron temperature at $T_e < 1.6\times 10^{4}$\,K and the oxygen abundance $12+\log{\rm (O/H)} > 7.75$. Instead, we derive the oxygen abundance using strong-line diagnostics of the detected nebular emission lines. We adopt a single set of calibrations from the literature for various line ratios, derived at $z=4-9$ with \jwst\ observations \citep{Sanders24}. To be conservative, we model the joint probability distribution function (PDF), where each applicable strong-line diagnostic is inversely weighted by the scatter in its respective relation (see Fig.~\ref{fig:metpdf}). The width of the individual PDFs includes both the statistical uncertainties from the line flux ratios and the systematic uncertainties from the line calibrations. The joint PDF yields a median $12+\log{\rm (O/H)} = 8.36 \pm 0.10$, where the uncertainty reflects the 16th to the 84th percentile of the distribution. We note that this method generally favors calibrations with the lowest scatter and minimal systematic uncertainties, such as O3 and R23, that are less affected by dust or flux calibration issues. Moreover, the inclusion of the O32 and O2 calibration breaks the degeneracy of the R23 and O3 diagnostics and suggests the upper-branch solution being correct for both.

\subsection{ALMA spatial and spectroscopic modeling.}

The observations and reduction of the ALMA [\cii]-$158\mu$m spectral cube covering A1689-zD1 have been presented previously \citep{Knudsen25}. For consistency, we perform an independent analysis of the [\cii]-$158\mu$m data cube here. We first derive the integrated intensity over the spectral line, the moment-0 map, shown in Figure~\ref{fig:jwstalma}. We model the spatial extent of the emission with a 2D Guassian function, and determine full-width-at-half-maximum (FWHM) major and minor axes of ${\rm FWHM}_{\rm maj},{\rm FWHM}_{\rm min} = 1.\!\!^{\prime\prime} 51, 0.\!\!^{\prime\prime} 79$. This corresponds to $2.6\times 1.3$\,kpc at $z=7.13$ in the source plane, corrected for lensing shear, $\sqrt{\mu}$. The [\cii]-$158\mu$m disk diameter is determined as $D_{\rm [CII]} = \sqrt{{\rm FWHM}_{\rm maj}\times {\rm FWHM}_{\rm min}} \times 1.5 = 2.8$\,kpc. The spatial profile reveals a fairly smooth, disk-like structure centered at the bright rest-frame UV/optical emission components. We note, however, that there do appear to be sub-structures in the high-resolution, high-S/N images presented by \citet{Knudsen25} which slightly complicate the analysis. Throughout this work, we focus on the overall, integrated properties of the galaxies, to also be able to compare more directly to literature values on similar massive, high-redshift galaxies.

The 1D spectrum is extracted from the full data cube by summing the flux over the entire [\cii] emitting region. The [\cii]-$158\mu$m transition is clearly detected ($>10\sigma$), and we derive a line FWHM of $366\pm 18$\,km\,s$^{-1}$. Integrating over the entire emission-line profile in velocity space yields a total line flux intensity of $3.9\pm 0.3$\,Jy\,km\,s$^{-1}$, which corresponds to a line luminosity \citep{Solomon05} of $L_{\rm [CII]} = (5.1\pm 0.4)\times 10^{8}\,L_\odot$ at the target redshift and corrected for the magnification factor \citep[consistent with that reported by][]{Knudsen25}.

%\clearpage
%\newpage

\section{The dust-to-metal ratios of galaxies in the epoch of reionization} \label{sec:dtmz}

\subsection{Low relative dust content in the metal-rich galaxy A1689-zD1 at $z=7.13$}

The physical properties derived from the comprehensive analysis of the \jwst\ and ALMA observations of A1689-zD1 are summarized in Table~\ref{tab:prop}. Based on the derived [\cii]-$158\mu$m line luminosity and the FWHM, we determine the total dynamical mass as $M_{\rm dyn}/\mathrm{M_\odot} = 1.16\times 10^{5}v_{\rm circ}^2 D_{\rm [CII]}$ \citep{Wang13}, where $v_{\rm circ}$ is the circular velocity and $D_{\rm [CII]}$ is the disk diameter in kpc calculated as the deconvolved Gaussian spatial FWHM$\times 1.5$ and corrected by the shear from the lens magnification, $\sqrt{\mu}$. We assume that the velocities are described by $v_{\rm circ} = 0.75\times {\rm FWHM}_{\rm [CII]} / \sin(i)$ \citep{Decarli18}, where $i=58^\circ$ is the approximate inclination angle measured as the ratio between the spatial [\cii] emission ellipsoidal minor and major axes, $i = \cos^{-1}({a_{\rm min}} / {a_{\rm maj}})$. This yields a dynamical mass of $M_{\rm dyn} = (3.1\pm 0.3)\times 10^{10}\,M_\odot$. We caution that this estimate might be biased if A1689-zD1 is actively merging. Generally, the dynamical mass provides an upper bound on the available gas mass within the [\cii]-emitting region of the galaxy for virialized structures \citep[e.g.,][]{DDessaugesZavadsky20,Rizzo21,Heintz22,Aravena24}. We complementarily infer the gas mass using the metallicity-dependent [\cii]-to-$M_{\rm gas}$ relation from \citet{Heintz21}, where $M_{\rm gas}/L_{\rm [CII]} \approx 31 \,M_\odot\,L^{-1}_\odot$ at solar metallicity \citep[consistent with the calibration by][]{Zanella18}. This relation has also later been reproduced by recent simulations \citep{Vizgan22,Casavecchia25}. This yields a gas mass of $M_{\rm gas} = (3.0\pm 0.7)\times 10^{10}\,M_\odot$ for A1689-zD1 given the derived [\cii]-$158\mu$m line luminosity and oxygen abundance $12+\log{\rm (O/H)}$. 

Subtracting the stellar component from the total dynamical mass implies a total gas mass of at most $M_{\rm gas} = (2.8^{+0.2}_{-1.7})\times 10^{10}\,M_\odot$, consistent with the gas mass inferred from the [\cii] line luminosity. The molecular gas mass is constrained to be $<7.2\times 10^{9}\,M_\odot$ based on the non-detection of the CO(3-2) transition \citep{Knudsen17}. This implies that the molecular gas fraction makes up a minor contribution to the total ISM budget, such that the inferred gas mass is likely mainly comprised of neutral, atomic \hi. In the following analysis, we assume the more conservative gas mass from the stellar-subtracted dynamical measure, where the reported error budget also conservatively accounts for the two independent approaches used to estimate the gas mass. We caution, however, that the dynamics of this system may be perturbed by a merger scenario \citep{Knudsen25} for example, such that the gas mass inferred from the dynamics could be higher than the real mass. We argue, however, this to be a minor effect given the consistency with the luminosity-inferred gas mass estimate.
%We therefore assume the gas mass derived from the line emission throughout this work. 

This measurement implies that the large majority of the baryonic matter in this galaxy is in the form of (neutral) gas. We determine a gas fraction $M_{\rm gas}/(M_\star + M_{\rm gas})\sim 0.9$, and a gas depletion time, $t_{\rm dep} = M_{\rm gas}/{\rm SFR_{\rm tot}} \sim 0.80$\,Gyr, where ${\rm SFR_{\rm tot}} = {\rm SFR_{SED}} + {\rm SFR_{IR}} = 34.8^{+5.0}_{-4.6}\,M_\odot\,{\rm yr}^{-1}$. This gas mass also yields a metal mass of $M_Z = (2.5^{+0.9}_{-1.6})\times 10^{8}\,M_\odot$, defined as $M_Z = M_{\rm gas} \times Z/Z_\odot \times Z_{\rm \odot,ref}$ with $Z_{\rm \odot, ref} = 0.018$ \citep{Asplund09}. The metal mass also implies a minimum stellar mass of order \(10^{10}\,M_\odot\) \citep{Peeples14}, consistent with the upper end of our uncertainty on the stellar mass from the SED fitting, and implying a gas mass fraction of about 0.7. These measurements combined yield dust-to-gas (DTG) and dust-to-metal (DTM) mass ratios for A1689-zD1 of $(5.5^{+3.5}_{-1.2})\times 10^{-4}$ and $(6.2^{+4.2}_{-2.6})\times 10^{-2}$, respectively, assuming the dust mass reported in Table~\ref{tab:prop}. The derived dust-to-stellar mass ratio is consistent with other similar estimates for galaxies at $z>6$ from the literature \citep[e.g.,][]{Witstok22,Algera25}, and implies a typical net dust yield $y_{\rm SN} \sim 0.13\,M_{\odot}$ per supernova \citep{Lesniewska19} but with an inefficient destruction in the reverse shock.

\begin{deluxetable}{lr}
\tablewidth{\linewidth}
\centering
\tabletypesize{\footnotesize}
\tablecaption{Physical properties of A1689-zD1. \label{tab:prop}}
\tablehead{
\colhead{Quantity} & \colhead{Value}
}
%\multicolumn{2}{c}{HST (ACS, WFC3)} \\
\hline \vspace{-0.18cm}\\
        R.A. (J2000) & $13^{\rm h}11^{\rm m}29^{\rm s}.92$ ($197^{\circ}.87467$) \\
        Decl. (J2000) & $-01^\circ 19^{\prime} 19.\!\!^{\prime\prime} 0$ ($-1^{\circ}.32194$) \\
        $z_{\rm spec}$ & $7.1324\pm 0.0001$ \\
                       \vspace{0.2cm}
        $M_{\rm UV}$ (mag) & $-17.8\pm 0.2$ \\
        SFR$_{\rm H\beta}$ ($M_\odot$\,yr$^{-1}$) & $3.9\pm 1.2$ \\
        12+log(O/H) & $8.36\pm 0.10$ \\
        $Z/Z_\odot$ & $0.47^{+0.12}_{-0.10}$ \\
        \vspace{0.2cm}
        $T_e$ (K) & $< 1.6\times 10^{4}$ ($3\sigma$) \\
            $M_\star$ $(10^{9}\,M_\odot)$ & $2.6^{+12.5}_{-0.6}$ \\
        $M_{\rm dust}$ $(10^{7}\,M_\odot)$ & $1.54\pm 0.32$ \\
        $T_{\rm dust}$ (K) & $43.0^{+4.7}_{-3.9}$ \\
        $A_V$ (mag) & $1.0\pm 0.3$\ \\
        SFR$_{\rm IR}$ ($M_\odot$\,yr$^{-1}$) & $27.1^{+4.4}_{-3.6}$ \\
        Mass-weighted age (Myr) & $41^{+52}_{-21}$ \\
        $M_{\rm gas}$ $(10^{10}\,M_\odot)$ & $2.8^{+0.2}_{-1.7}$ \\
        $M_{Z}$ $(10^{8}\,M_\odot)$ & $2.5^{+0.9}_{-1.6}$ \\
        \vspace{0.2cm}
        $M_{\rm dyn,tot}$ $(10^{10}\,M_\odot)$ & $3.1\pm 0.3$ \\
        DTG mass ratio & $(5.5^{+3.5}_{-1.2})\times 10^{-4}$ \\
        DTM mass ratio & $(6.2^{+4.2}_{-2.6})\times 10^{-2}$
\enddata
\tablecomments{All quantities reported here that are affected by magnification due to gravitational lensing have been corrected by the magnification factor $\mu = 9.6\pm 0.2$. Uncertainties are quoted at $1\sigma$ confidence, and upper limits at $3\sigma$.}
\end{deluxetable}

\subsection{Evolution of DTG and DTM mass ratios with metallicity at $z>6$} \label{ssec:dtmmet}

\begin{figure}[!t]
\centering
\includegraphics[width=9cm]{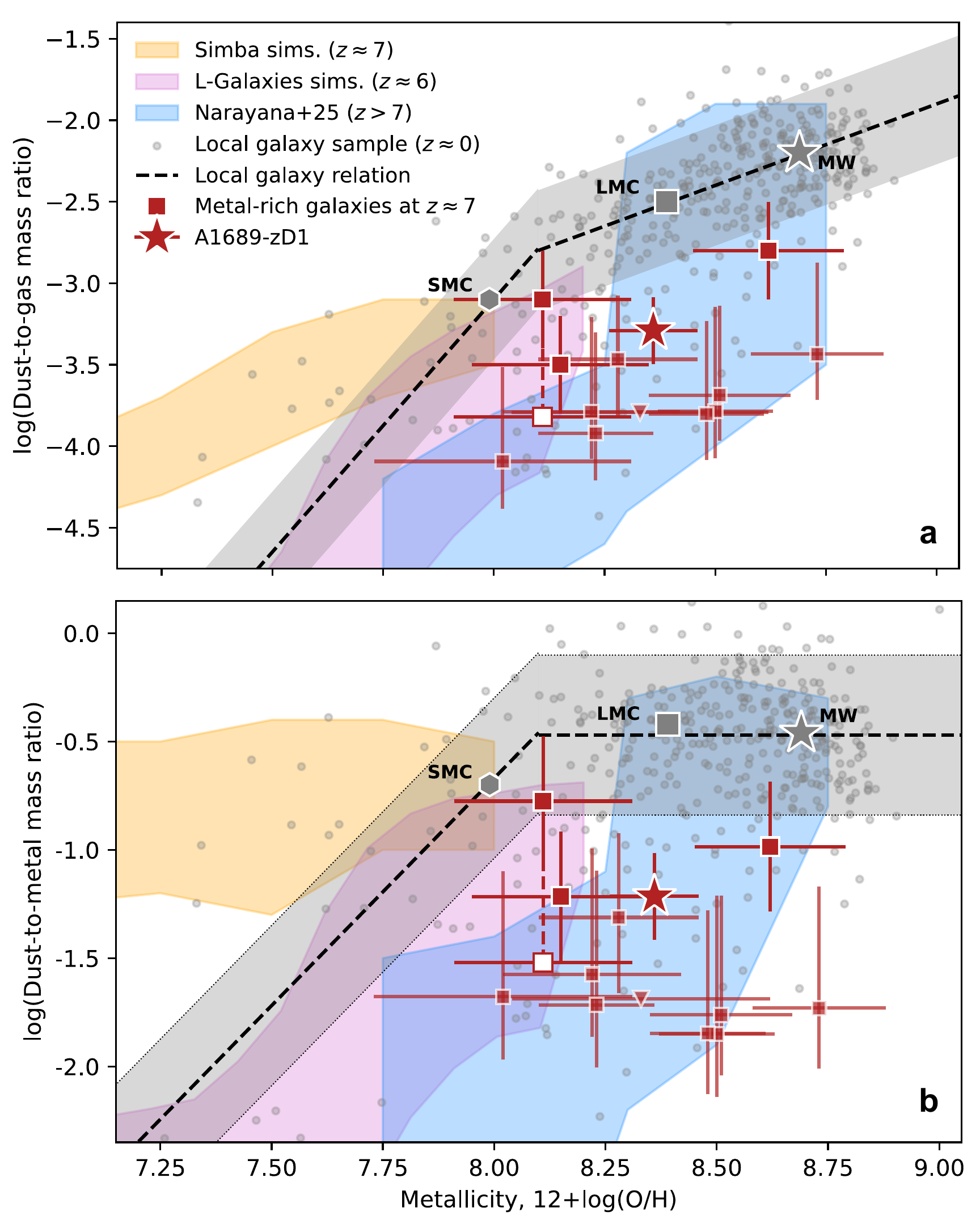}
\caption{\textbf{Relative dust mass as a function of metallicity, 12+log(O/H).} Panel (a): The dust-to-gas mass ratio of A1689-zD1 (red star symbol) and the other compiled metal-rich galaxies at $z>6$ (red squares) compared with a sample of local galaxies \citep{RemyRuyer14,DeVis19}, the empirical, metallicity-dependent relation for local galaxies (dashed black line, light-shaded area indicate $1\sigma$) \citep{RemyRuyer14} and predictions from simulations at $z\gtrsim 6-7$ \citep[yellow, purple, and blue regions denote $1\sigma$ scatter;][]{Dave19,Yates24,Narayanan25}. The potential lower relative dust mass of the galaxy z6.3 is shown by the white square, connected by the red dashed line. Errorbars show $1\sigma$ confidence intervals. The average values for the Milky Way (MW) and the Large and Small Magellanic Clouds (LMC and SMC) are shown as the grey star, square, and hexagon symbols \citep{Konstantopoulou24}, respectively. Panel (b): The dust-to-metal mass ratio as a function of metallicity, 12+log(O/H), following the same symbol notation as Panel~(a). }
\label{fig:dtm}
\end{figure}

% Note here on similarity to JWST-only measurement.
To place A1689-zD1 in context, we compare in Figure~\ref{fig:dtm} the DTG and DTM mass ratios and metallicity of A1689-zD1 to samples of galaxies in the local Universe \citep{RemyRuyer14,DeVis19}, together with empirical scaling relations at redshifts $z\approx 0$, expectations from cosmological simulations at $z\approx 6-7$ \citep{Dave19,Yates24,Narayanan25}, and the average values for the Milky Way and the Large and Small Magellanic Clouds \citep{Konstantopoulou24}. A1689-zD1 has an order of magnitude lower DTG and DTM mass ratio (at $>3\sigma$) compared to the Milky Way and other local galaxies \citep{DeVis19,Galliano21} with similar chemical enrichment ($12+\log{\rm O/H} > 8.3$), and to estimates of star-forming galaxies at redshifts $z\approx 2-3$ \citep{Shapley20}. We also calculate the DTG and DTM for three other metal-rich galaxies at redshifts $z>6$ that have similar measurements to A1689-zD1, with multiple FIR continuum detections to robustly constrain the dust temperature and mass and dynamical [\cii]-$158\mu$m emission measurements to constrain the total gas mass; the lensed z6.3 galaxy or ``Cosmic Grapes'' at $z=6.072$ \citep{Fujimoto25_z63,Valentino24,GimenezArteaga24}, REBELS-25 at $z=7.306$ \citep{Bouwens22,Rowland24,Algera25}, and
B14-65666 at $z=7.15$ also known as the ``Big Three Dragons'' \citep{Hashimoto19,Sugahara25,Jones24}. These are all characterized by relatively high oxygen abundances, $12+\log{\rm (O/H)}> 8.0$. Furthermore, we include data on the dozen galaxies from the REBELS survey \citep{Algera25}, which provides ALMA and \emph{JWST} measurements for bright galaxies at \(z>6\) (though most only with a single FIR continuum measurement). We use, however, a more typical infrared slope, $\beta_{\rm IR}=1.7$ \citep{Witstok22} and a metallicity-dependent [\cii]-to-$M_{\rm gas}$ scaling for a more direct comparison to A1689-zD1 in the plot. 

The high-redshift sources generally show an order of magnitude lower DTG and DTM mass ratios (combined at $>3\sigma$) than local galaxies with a similar chemical enrichment. This result is still significant if we use gas masses inferred from the stellar mass--subtracted dynamical mass, or a constant [\cii]-to-$M_{\rm gas}$ conversion factor based on the [\cii]-to-H\(_2\) relation \citep{Zanella18} as tracer of the total gas mass in the galaxies. It is notable that local galaxies with equivalent DTG or DTM mass ratios to the galaxies at \(z>6\) are mostly found with metal abundances $\sim 10\%$ of the solar value \citep{DeVis19}. We also emphasize that this order-of-magnitude lower DTG mass ratio is consistent with the ratio of the visual extinction to the \hi\ gas column, $A_V/N_{\rm HI}$ for A1689-zD1, which again is more than an order of magnitude lower than the Galactic average and predictions for the metallicity-evolution of the DTM at $z<6$ from $\gamma$-ray burst sightlines. The consistently low dust-to-gas and dust-to-metal ratios observed for A1689-zD1, derived through two independent methods, are great validations of the individual approaches and the overall result for this source. We caution here that this population of galaxies are partly selected for ALMA follow-up based on their UV brightness, and that fully optically obscured sources \citep[e.g.,][]{Fudamoto21} might show higher relative dust masses.

\begin{figure*}[!t]
\centering
\includegraphics[width=15cm]{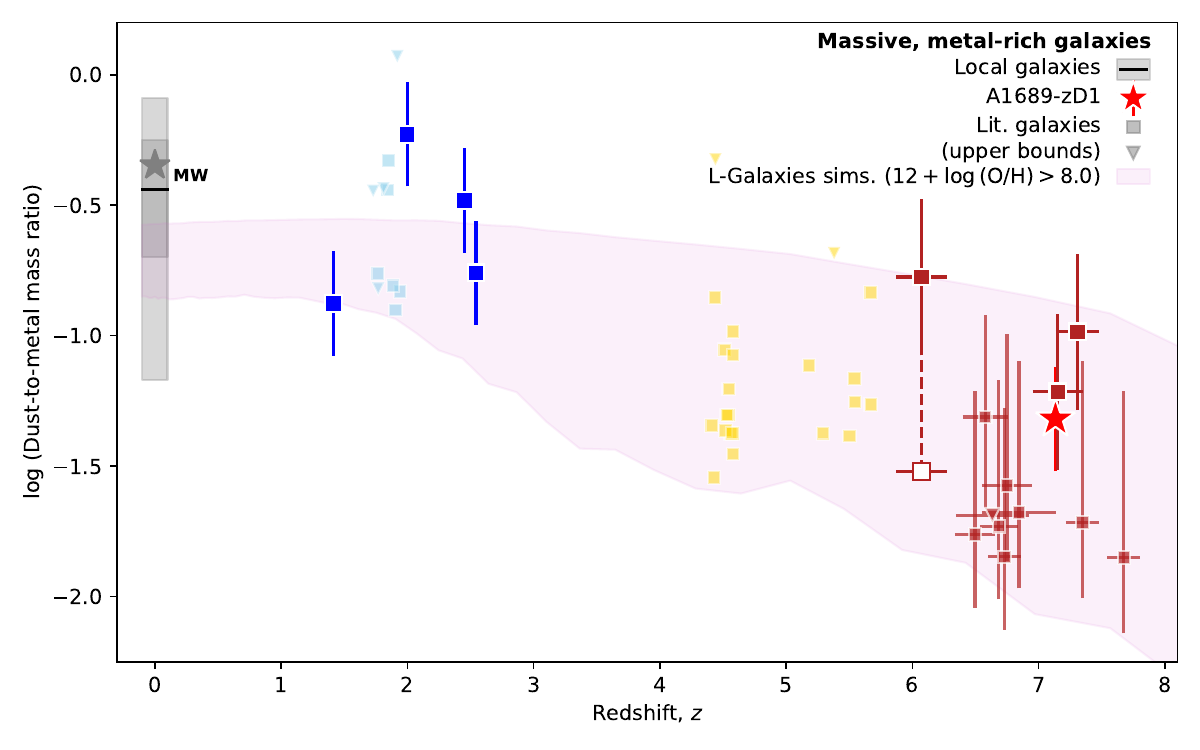}
\caption{\textbf{Dust-to-metal mass ratio of massive, metal-rich galaxies as a function of redshift, $z$.} The primary source A1689-zD1 is shown as the red star symbol. The Galactic average is shown as the grey star symbol. The median (black solid line) and $1\sigma$ and $2\sigma$ sample interval (dark- and light area) are shown for a local sample of metal-rich ($12+\log{\rm (O/H) > 8.0}$) galaxies \citep{DeVis19}. The predicted DTM evolution for metal-rich ($12+\log{\rm (O/H)}>8.0$), star-forming galaxies from the {\tt L-Galaxies} simulations is highlighted by the purple-shaded region denoting the $1\sigma$ percentile of the full distribution \citep{Yates24}. Highlighted blue, yellow, and red squares mark metal-rich galaxies at $z=2-3$, $z\approx 5$, and $z\approx 7$ from the literature with dust and metal mass estimates. The $z>6$ symbol notation follow Fig.~\ref{fig:dtm}. Light-shaded squares (triangles) highlight dust-to-metal mass measurements (upper bounds) for samples of high-redshift galaxies using a proxy for the metal mass.}
\label{fig:dtmz}
\end{figure*}

% Notes on simulations here. 
Galaxy evolution simulations suggest that a potential flattening of the DTG mass ratio below $M_{\rm dust}/M_{\rm gas} \sim 10^{-3}$ at $12+\log{\rm (O/H)} > 8.0$ is expected to occur in high-redshift galaxies when in-situ grain growth starts to become efficient and is roughly balanced by dust destruction from supernova shocks \citep{Yates24}. Indeed, we find that the lower DTG and DTM mass ratios of the compiled galaxies at $z>6$ are consistent with the extrapolated flattening predicted from the {\sc Simba} and {\sc L-Galaxies} simulations in Fig.~\ref{fig:dtm}. Unfortunately neither of these include the metal-rich environments of the galaxies observed here, excluding a direct comparison. The recent cosmological zoom-in simulations from \citet{Narayanan25}, which uses an updated dust evolution scheme \citep[see][for details]{Narayanan23}, accurately predict the lower DTG and DTM mass ratios of galaxies at $z>7$. However, these simulations show an upturn at $12+\log{\rm (O/H)} > 8.3$ for galaxies at $z\lesssim 10$, corresponding to the timescales for which the dust transitions being primary from stellar production to dominated by ISM growth. The observed DTG and DTM mass ratios do not show a similar upturn, which might indicate that this process likely happens later, at $z<6$, potentially related to the typical star-formation histories of the galaxies. This conclusion is, however, limited by the sparse numbers of the most metal-rich galaxies.   
% We note that the observed mass ratios require future simulations to maximally produce DTM mass ratios of $\log {\rm DTM}\approx -1.5$ even for near-solar metallicity environments at $z>6$. 
These results warrant a more detailed look into the redshift evolution of in particular the DTM mass ratio, since dust grains are expected to coagulate more efficiently in more metal-rich interstellar environments, causing a correlation between the DTM and the metallicity of the gas \citep{Mattsson14}. 
%which should be more directly linked to the dust grain properties and growth.

\subsection{Evolution of the DTM mass ratio with redshift}

To investigate more directly the dust production from the available metals as a function of redshift, we find that massive galaxies at redshifts $z\gtrsim 4$ show systematically lower relative dust masses than at later times, see Figure~\ref{fig:dtmz}. This indicates a clear change in the amount or composition of cosmic dust in chemically enriched galaxies early in the history of the universe compared to their later counterparts.  It is important to note that this effect is physically distinct from the decrease in dust content with increasing redshift observed in quasar or $\gamma$-ray burst sightlines \citep{Peroux20,Heintz23_GRB}, which is primarily a representation of the lower metallicities in these systems at early times. The observations are overall in good qualitative agreement with predictions for metal-rich ($12+\log{\rm (O/H)}>8.0$), star-forming galaxies from the {\tt L-Galaxies} simulations \citep{Yates24}, predicting a gradual increase in the DTM with decreasing redshift as shown in Fig.~\ref{fig:dtmz}. Comparatively, the simulations from \citet{Narayanan25} instead show a rapid upturn in the DTM as discussed in Sect.~\ref{ssec:dtmmet} which might resemble the observations better, though predicted to happen at $z\sim 10$. Delaying their transition for dust buildup from mainly stellar production to ISM growth by $\sim 500$\,Myr remarkably well reproduces the full set of observations from $z=0-8$. These new ALMA+\jwst\ observations are also qualitatively in agreement with the early results reported by \citet{Capak15}, showing a clear trend of galaxies at $z\gtrsim 5$ with lower dust masses to [\cii] line luminosities (tracing the total gas or metal mass) relative to their local counterparts. 

There are several possible explanations for this effect. Models that predict lower DTM in the early universe \citep{Yates24,Narayanan25} seem to do so because the efficiency of grain growth in the interstellar medium does not keep pace with the production of metals in core collapse supernovae because of the high star-formation rates, or rather the short gas depletion times of high-\(z\) galaxies \citep{Feldmann15}. This is particularly true for low-mass galaxies, but is also true to a lesser extent for massive galaxies such as A1689-zD1 as their dust reservoirs builds up. Curiously, the dust-to-stellar mass ratio for A1689-zD1 is consistent with dust production solely by core-collapse supernovae, i.e.\ net production of around 0.1\,M\(_\odot\) per supernova \citep{Gall18,Kirchschlager24}, which may point to SN-dominated dust production and much less interstellar grain growth. Another possibility is that asymptotic giant branch (AGB) stars matter significantly more in terms of dust production (either in term of total mass of the production of high-emissivity dust), since the age of the universe at \(z\sim4-6\) corresponds to the age of the most massive AGB stars. Finally, high-redshift galaxies may produce the same amount of dust, but are more hostile environments for dust grains, either preventing growth on rapid timescales, or destroying dust efficiently \citep[e.g.,][]{Ferrara16}.

Alternatively, the difference may not be the dust mass, but rather the FIR emissivity of the grains. It may be that a low-emissivity effect is occurring in the early universe instead of genuinely low DTG and DTM. This is intriguing in the context of the now mounting evidence for a remarkable overabundance of UV-bright galaxies in the early universe compared to theoretical and past empirical expectations \citep[e.g.,][]{Harikane24}; a promising scenario to explain this discrepancy is as a result of lower dust attenuation, either due to varying radiative transfer or star-dust geometry effects \citep{Ferrara23} or a preferential production of large dust grains \citep{Toyouchi25,Narayanan25}, which may be consistent with the drop in DTM we report here. Additionally, we may be witnessing galaxies that rapidly reach high metal abundances during their early evolutionary stages if they host a more top-heavy IMF \citep{Eales24}, effectively diluting their DTM values. We refer to \citet{Sommovigo26} for a more comprehensive review on the physical conditions and impact of varying geometry and grain size distribution on the observed galaxy properties.

In general, there has been various recent work using novel simulations to describe the dust content of early galaxies. Previous simulations seem to reproduce the low dust-to-gas and dust-to-metal mass ratios observed here, however only at low metallicities \citep[$\lesssim 20\%$ solar;][]{Popping17,Vijayan19,DiCesare23,Mauerhofer23}. This has been explained by reduced accretion efficiencies caused by a combination of low galactic metallicities and extremely bursty star formation \citep[][but see the discussion relating to the \textsc{L-Galaxies} simulation above]{Choban25}. The predicted dust mass estimates from the simulations are generally consistent with the observed dust-to-stellar mass ratios of high-redshift galaxies \citep{DiCesare23}. However, as hinted at in \citet{Algera25} and supported by the findings provided here, the gas and metal abundance of these early massive galaxies provide the necessary environments to in principle sustain an even more efficient grain growth and higher dust abundance, which are not observed. This suggests that metals can be produced rapidly from the first massive stars formed, but dust production is halted. This is likely either due to the timescales involved in their aggregation or coagulation, or from more efficient destruction mechanisms in the more turbulent, ionized media of early massive galaxies. In the \textsc{L-Galaxies} simulations, dust is destroyed in the CGM through sputtering, with a slight impact on the cycling of dust out and back into galaxies. However, this CGM sputtering only occurs at a rate comparable to galactic dust processes in massive early-type galaxies at lower redshift.

\section{Summary \& Outlook}
\label{sec:conc}

%Summary and Future outlook.
In this work, we performed a comprehensive, multi-wavelength analysis of new and publicly available observations with \jwst\ and ALMA of the bright, highly lensed galaxy A1689-zD1 at $z=7.1332$. This galaxy has since its discovery held the status as the prototype for dusty `normal' galaxies during the epoch of reionization at $z>6$. We revisited this galaxy to gauge the baryonic matter components in the ISM, with particular focus on constraining the build up of cosmic dust. We performed rest-frame UV to FIR modeling of the SED of A1689-zD1 using multiple SED fitting codes for validation checks, to determine its stellar mass, dust mass, visual attenuation, and star-formation rate. The ALMA observations were also used to constrain the total dynamical mass of the source, and infer the gas mass using common gas tracers but bounded by the overall dynamics of the system. We further analyzed new \jwst/NIRSpec medium-resolution IFU observations to measure the line luminosities of the most prominent rest-frame optical nebular emission lines seen in the spectra, with particular focus on determining the gas-phase metallicity, quantified via the oxygen abundance $12+\log{\rm (O/H)}$. 

Using the rest-frame UV/optical SED model as the intrinsic galaxy template, we modeled the remarkably strong damped Lyman-$\alpha$ absorption feature seen in the VLT/X-shooter spectrum \citep{Watson15} directly from the photometry, yielding a high \hi\ column density in the line-of-sight of $N_{\rm HI}\sim 10^{22.8}\,$cm$^{-2}$. From this we determined a relatively low dust-to-gas ratio of $A_V/N_{\rm HI} = (1.3^{+2.7}_{-0.9})\times 10^{-23}$\,mag\,cm$^{-2}$, more than an order of magnitude lower than that found in the Milky Way and the LMC and SMC and an empirical metallicity-dependent DTG relation for $\gamma$-ray burst host galaxies at $z=1.7-6.3$ at the derived metallicity $12+\log({\rm O/H}) = 8.36\pm 0.10$ of A1689-zD1. This suggests that the bulk \hi\ gas in the line-of-sight to A1689-zD1 is relatively dust-poor compared to its chemical enrichment. 

The derived dust-to-stellar mass ratio implied a typical net dust yield $y_{\rm SN} \sim 0.13\,M_{\odot}$ per supernova, consistent with other contemporary measurements. 
However, based on the substantial gas mass, $M_{\rm gas} = (2.8^{+0.2}_{-1.7})\times 10^{10}\,M_\odot$, and the total metal mass of the ISM, $M_Z = (2.5^{+0.9}_{-1.6})\times 10^{8}\,M_\odot$, we inferred, the DTG and DTM mass ratios were found to be remarkably low, with ${\rm DTG} = (5.5^{+3.5}_{-1.2})\times 10^{-4}$ and ${\rm DTM} = (6.2^{+4.2}_{-2.6})\times 10^{-2}$, respectively. These are roughly an order of magnitude lower than the typical values found for galaxies with similar chemical enrichment in the local Universe, consistent with the \jwst-only measurement of the DTG ratio quantified via $A_V/N_{\rm HI}$. These two independent measurements corroborate our results for A1689-zD1 and indicate a potential change in the relative dust abundance or composition of early galaxies. 

To quantify this, we compiled a set of galaxies at $z>6$ jointly observed with \jwst\ and ALMA from the literature, for which similar constraints on the DTG and DTM mass ratios were attainable. We found that all metal-rich galaxies at these redshifts with $12+\log({\rm O/H}) > 8.0$ show a similar deficit in their relative dust masses, suggesting a universal transition in the dust properties of early galaxies despite their otherwise chemically matured environments. We theorized that this was likely due to a change in the primary dust production channels, more efficient destruction mechanisms, or a general change in the emissivity or attenuation properties of the dust grains due to a different dust grain-size distribution and composition. 
%Note on theories (new Simba sims).

Our observations hint at a remarkable transition in the properties or visibility of dust grains in the early Universe, even in chemically mature environments. This is particularly intriguing in the context of the observed overabundance of UV-bright galaxies at $z\gtrsim 10$ \citep[e.g.,][]{Harikane24}. One of the most promising theories suggests lower dust abundances or attenuation to explain their higher UV luminosities \citep{Ferrara23,Toyouchi25,Narayanan25}, although the exact redshift for the transition is still debated (but likely at $z\gtrsim 8$). The analysis presented here provides empirical support for this scenario, though suggests that the transition might occur later in the history of the Universe at $z\sim 3-4$, roughly 1.5\,Gyr after the Big Bang. This timescale corresponds to the age of the most massive AGB stars, such that these sources might start to dominate the dust production, significantly altering the abundances, composition and attenuation profiles of cosmic dust in the ISM of galaxies beyond this epoch. Dedicated joint \jwst\ and ALMA observations in the future are key to fully chart this evolution to constrain the evolution of the primary dust ingredients at high redshifts.

\section*{Acknowledgments}

The Cosmic Dawn Center (DAWN) is funded by the Danish National Research Foundation under grant DNRF140. Some of the data products presented herein were retrieved from the Dawn \jwst\ Archive (DJA). DJA is an initiative of the Cosmic Dawn Center (DAWN), which is funded by the Danish National Research Foundation under grant DNRF140. KEH acknowledges support from the Independent Research Fund Denmark (DFF) under grant 5251-00009B and co-funding by the European Union (ERC, HEAVYMETAL, 101071865). Views and opinions expressed are, however, those of the authors only and do not necessarily reflect those of the European Union or the European Research Council. Neither the European Union nor the granting authority can be held responsible for them.  J.R.W. acknowledges that support for this work was provided by The Brinson Foundation through a Brinson Prize Fellowship grant. K.K. acknowledges support from the Knut and Alice Wallenberg Foundation (KAW 2020.0081 and KAW 2023.0339) as well as support from the ERC synergy grant 101166930 (RECAP).

Software used in developing this work includes: \texttt{matplotlib} \citep{matplotlib}, \texttt{numpy} \citep{numpy}, \texttt{scipy} \citep{scipy}, and \texttt{Astropy} \citep{Astropy}.

This work is based on observations made with the NASA/ESA/CSA James Webb Space Telescope. The data were obtained from the Mikulski Archive for Space Telescopes at the Space Telescope Science Institute, which is operated by the Association of Universities for Research in Astronomy, Inc., under NASA contract NAS 5-03127 for \textit{JWST}.

\bibliography{ref}
\bibliographystyle{apj}

%\begin{appendix}
%\section{If needed}

%\end{appendix}

\end{document}